\documentclass[aps,preprint,tightenlines,showpacs,showkeys]{revtex4-1}
\usepackage[latin9]{inputenc}
\usepackage{amsmath}
\usepackage{amssymb}

\makeatletter

\usepackage{dcolumn}
\usepackage{bm}

\makeatother

\begin{document}

\title{$D=4$ topological gravity from gauging the Maxwell-special-affine
group}

\author{Salih Kibaro\u{g}lu$^{1}$\thanks{E-mail: salihkibaroglu@gmail.com},
Mustafa \c{S}enay$^{1}$\thanks{E-mail: mustafasenay86@hotmail.com}
and Oktay Cebecio\u{g}lu$^{2}$ \thanks{E-mail: ocebecioglu@kocaeli.edu.tr}}

\date{\today}
\begin{abstract}
In this paper, the Maxwell extension of the special-affine algebra
is obtained and corresponding non-linear realization is constructed.
We give also the differential realization of the generators of the
extended symmetry. Moreover, we present the gauge theory of the Maxwell
special-affine algebra and the topological gravity action in four
dimensions. As a conclusion, we show that the Bianchi identities can
be found by using the solution of the equations of motion.
\end{abstract}

\affiliation{$^{1}$Naval Academy, National Defence University, 34940 \.{I}stanbul,
Turkey }

\affiliation{$^{2}$Department of Physics, Kocaeli University, 41380 Kocaeli,
Turkey}
\maketitle

\section{Introduction}

There has been a great deal of interest in the gauge theoretical description of gravitation, which was originally proposed by Utiyama, Kibble and Sciama \cite{utiyama1956,kibble1961,sciama1964}. For instance, the Poincar\'{e} gauge theory of gravity is a gauge theory based on Poincar\'{e} group that generalizes the Riemann geometry to include torsion in addition to curvature \cite{kibble1961}. Likewise, various extended gravity theories are obtained by using gauge theories of different Lie groups \cite{macdowell1977,kaku1977,charap1974}. 

The affine group $A(4,R)$ is the semi-direct product of the general linear group $GL(4,R)$ and the translations, so it can be seen as an extension of the Poincar\'{e} group where the Lorentz group is replaced by $GL(4,R)$. In comparison with Lorentz group, the $GL(4, R)$ contains both volume changing dilation and volume preserving shear transformations. In 1974 Yang \cite{yang1974} put forwarded a gauge theory of gravity based on the affine group to construct a theory of (quantum) gravity in the high energy limit without any reference to supersymmetry or extra dimensions \cite{hehl1989}. Another study, named as the metric-affine gauge theory of gravity (MAG), generalized the Poincar\'{e} gauge theory of gravity with non-vanishing nonmetricity tensor\cite{hehl1976,hehl1995}. Later on, the papers \cite{neeman1979A,neeman1979B,neemann1988A,neemann1988B,lopez-pinto1995} suggested that the renormalizability and unitarity problems in quantum gravity can be solved by taking the affine group as the dynamical group in a gauge theory of gravity by the help of general linear connection $\Gamma^{\alpha}_{\mu\nu}$\cite{hehl1995}.

There is an interesting study which is based on $GL(4,R)$ gauge theory of gravity, proposes a unified theory between the electromagnetic and gravitation fields in the concept of purely affine formulation of the Einstein-Maxwell theory \cite{ferraris1982}. The purely affine formulation of gravity describes gravitational Lagrangian density in terms of a torsionless affine connection and the symmetric part of the Ricci tensor of the connection (a short review can be found in \cite{poplawski2008}). Likewise, Pop\l awski suggests a combination between the electromagnetic field and cosmological constant in the purely affine formulation\cite{poplawski2009}.

The special linear group $SL(n,R)$ is a subgroup of $GL(n,R)$ and contains Lorentz and shear transformations. In particle physics, it is used for group theoretical classification of hadrons \cite{gronwald1997} and plays an important role to describe the observed sequences of angular momentum excitation of hadrons \cite{hehl1977}. Moreover the observed structures of Regge trajectories can be explained by the group $SL(3,R)$ of deformations of hadronic matter \cite{ dothan1965,lord1978}. The fundamental importance of $SL(3,R)$ transformations for hadronic matter is reviewed in \cite{hehl1978}.

After these preliminaries and taking account of the developing of
gauge theory, we can easily say the extension of the affine group
can provide us new symmetries, in other words, new interactions. In
the light of this idea, we consider the Maxwell extension of the special-affine group $SA\left(4,R\right)$. The Maxwell symmetry is based on a noncentral extension of the Poincar\'{e} group which contains six new additional tensorial abelian generators $Z_{ab}=-Z_{ba}$ and for which the momentum operators satisfy the relation $\left[P_{a},P_{b}\right]=iZ_{ab}$ \cite{bacry1970,schrader1972, soroka2005}. Moreover, the Maxwell symmetries extend the Minkowski space with new background field. If we take this background field as an electromagnetic (e.m.) field  \cite{bacry1970,schrader1972}, the motion of relativistic particle in a constant e.m. field can be described by the Maxwell symmetries. After Soroka's paper \cite{soroka2005}, the Maxwell symmetry become popular, and is now
studied in several different fields of physics \cite{soroka2005,gomis2009,bonanos2009,bonanos2010A,bonanos2010B,bonanos2010C,azcaraga2011,durka2011,soroka2012,durka2012,azcarraga2013A,fedoruk2013,azcarraga2014,cebecioglu2014,cebecioglu2015},
especially in the field of gauge
theories of gravity and the explanation of cosmological constant problem
\cite{azcaraga2011,soroka2012,azcarraga2013A,cebecioglu2014}. In
our previous study \cite{cebecioglu2015} we extend the Maxwell group $\mathcal{M}\left(4,R\right)$
to the Maxwell-affine group $\mathcal{MA}\left(4,R\right)$ and discussed its gauge theory of gravity.

The present paper is organized as follows. In Section 2, we summarize our previous
work \cite{cebecioglu2015}. In Section 3, we construct the Maxwell-special-affine
group $\mathcal{M}\mathcal{SA}\left(4,R\right)$, we also present the non-linear realization of $\mathcal{M}\mathcal{SA}\left(4,R\right)$
to get the transformation rules of the generalized coordinates and
the differential realization of generators. In Section 4, we find gauge transformation rules of the gauge connections and curvatures. We also give the Bianchi identities. In Section 5, we propose a gravity action by using topological terms. In addition, the equations
of motion of related action are found. Section 6 concludes the letter.

\section{Maxwell-affine algebra and the gravity action}

The affine group $A\left(4,R\right)$ is defined as the group of all
linear transformations in 4D space. The affine group is a semi direct
product of $GL\left(4,R\right)$ and translation group $T\left(4\right)$
\cite{borisov1974},
\begin{equation}
x^{a}=\Lambda_{\,\,b}^{a}x^{b}+c^{a},
\end{equation}
where $\Lambda^{a}_{b}$ and $c^{a}$ represents linear transformations and translation respectively. In this chapter, we will summarize our previous
work \cite{cebecioglu2015} which contains an extension of the affine
group named as the Maxwell-Affine group $\mathcal{MA}\left(4,R\right)$
and its gravity action. The non-zero commutation relationships of
$\mathcal{MA}\left(4,R\right)$ in four dimensions are given, 
\begin{eqnarray}
\left[L_{\,\,b}^{a},L_{\,\,d}^{c}\right] & = & i\left(\delta_{\,\,b}^{c}L_{\,\,d}^{a}-\delta{}_{\,\,d}^{a}L{}_{\,\,b}^{c}\right),\nonumber \\
\left[L_{\,\,b}^{a},P_{c}\right] & = & -i\delta_{\,\,c}^{a}P_{b},\nonumber \\
\left[P_{a},P_{b}\right] & = & iZ_{ab},\nonumber \\
\left[L_{\,\,b}^{a},Z_{cd}\right] & = & i\left(\delta_{\,\,d}^{a}Z_{bc}-\delta_{\,\,c}^{a}Z_{bd}\right),
\end{eqnarray}
where the small Latin indices $a,b,...=0,...,3$ and the generators
$P_{a}$, $L_{\,\,b}^{a}$ and $Z_{ab}$ corresponds to translation
symmetry, the transformations of $GL\left(4,R\right)$ and the Maxwell
symmetry which contains six additional tensorial charges that behave
as a tensor under Lorentz transformations.

The Maurer-Cartan (MC) 1-forms which is defined as $\Omega=-ig^{-1}dg$,
here $g$ is the general element of the $\mathcal{MA}\left(4,R\right)$
and the structure equation is given as,
\begin{equation}
d\Omega+\frac{i}{2}\left[\Omega,\Omega\right]=0.\label{eq: MC eq}
\end{equation}
Thus, one can show that the MC 1-forms satisfy following equations,
\begin{eqnarray}
0 & = & d\Omega_{\,\,\,P}^{a}+\Omega_{\,\,Lb}^{a}\wedge\Omega_{\,\,\,P}^{a},\nonumber \\
0 & = & d\Omega_{\,\,Lb}^{a}+\Omega_{\,\,Lc}^{a}\wedge\Omega_{\,\,Lb}^{c},\nonumber \\
0 & = & d\Omega_{\,\,\,Z}^{ab}+\Omega_{\,\,Lc}^{[a|}\wedge\Omega_{\,\,\,Z}^{c|b]}-\frac{1}{2}\Omega_{\,\,\,P}^{a}\wedge\Omega_{\,\,\,P}^{b},
\end{eqnarray}
where the subscripts $P, L, Z$ represents generator labels and the antisymmetrization of tensors is defined by $A^{[a}B^{b]}=A^{a}B^{b}-A^{b}B^{a}$
throughout the paper. Using the gauge field $\mathcal{A}=e^{a}P_{a}+B^{ab}Z_{ab}+\tilde{\omega}_{\,\,a}^{b}L_{\,\,b}^{a}$,
the curvature 2-form $\mathcal{\digamma}$ can be found by the structure equation,
\begin{equation}
\mathcal{\digamma}=d\mathcal{A}+\frac{i}{2}\left[\mathcal{A},\mathcal{A}\right],
\end{equation}
and putting the definition $\mathcal{\digamma}=F^{a}P_{a}+F^{ab}Z_{ab}+\widetilde{R}{}_{\,\,\,a}^{b}L{}_{\,\,\,b}^{a}$
on the last equation we find 2-form curvatures as,
\begin{eqnarray}
F^{a} & = & de^{a}+\widetilde{\omega}{}_{\,\,\,b}^{a}\wedge e^{b},\nonumber \\
F^{ab} & = & dB^{ab}+\widetilde{\omega}{}_{\,\,\,\,\,\,c}^{[a|}\wedge B^{c|b]}-\frac{1}{2}e^{a}\wedge e^{b},\nonumber \\
\widetilde{R}{}_{\,\,\,b}^{a} & = & d\widetilde{\omega}{}_{\,\,\,b}^{a}+\widetilde{\omega}{}_{\,\,\,c}^{a}\wedge\widetilde{\omega}{}_{\,\,\,b}^{c},\label{eq: curv1}
\end{eqnarray}
where $e^{a}\left(x\right)$, $B^{ab}\left(x\right)$ and $\widetilde{\omega}_{\,\,\,b}^{a}\left(x\right)$
are related gauge fields. Considering the affine exterior covariant
derivative defined as $\mathcal{D}=d+\widetilde{\omega}$, thus
the Bianchi identities can be given as follows,
\begin{eqnarray}
\mathcal{D}F^{ab} & = & \widetilde{R}_{\,\,\,c}^{[a|}\wedge B^{c|b]}-\frac{1}{2}F^{[a}\wedge e^{b]},\nonumber \\
\mathcal{D}F^{a} & = & \widetilde{R}{}_{\,\,\,b}^{a}\wedge e^{b},\nonumber \\
\mathcal{D}\widetilde{R}{}_{\,\,\,b}^{a} & = & 0.
\end{eqnarray}

Using the curvatures (\ref{eq: curv1}), the definition of shifted
curvature $\mathcal{Y}{}_{\,\,\,b}^{a}:=\widetilde{R}{}_{\,\,\,b}^{a}-\mu F{}_{\,\,\,b}^{a}$
which transforms covariantly under $GL(4,R)$ symmetry, and Pontryagin
densities in four dimensions \cite{mardones1991}, one can write a gauge
invariant Lagrangian as,
\begin{equation}
S=\frac{1}{2\varkappa}\int\mathcal{Y}{}_{\,\,\,b}^{a}\wedge\mathcal{Y}{}_{\,\,\,a}^{b}+\frac{1}{\rho}\int F^{a}\wedge F_{a},\label{action-1-1}
\end{equation}
where $\varkappa$ and $\rho$ are constant. By construction it is
easy to show that the action is diffeomorphism invariant and satisfies
local $GL(4,R)$ invariance \cite{cebecioglu2015}.

\section{Special-affine algebra and its Maxwell extension}
The $GL\left(4,R\right)$ group can be split into the one parameter group of dilations, and the $SL\left(4,R\right)$ group which contains volume preserving shear transformations in the Minkowski space-time. This provide us a restricted framework in which dilation transformation is broken comparing with our previous study \cite{cebecioglu2015}. The special-linear
algebra is generated by following traceless generators \cite{mielke2011,mielke2012},
\begin{equation}
\mathring{L}_{\,\,b}^{a}=L_{\,\,b}^{a}-\frac{1}{4}\delta_{\,\,b}^{a}L_{\,\,c}^{c}.
\end{equation}

In order to obtain the special-affine group $SA\left(4,R\right)$
we take the semi-direct product of $SL\left(4,R\right)$ and translational
symmetry generated by $P_{a}$ and its Lie algebra is defined by,
\begin{eqnarray}
\left[\mathring{L}_{\,\,b}^{a},\mathring{L}_{\,\,d}^{c}\right] & = & i\left(\delta_{\,\,b}^{c}\mathring{L}_{\,\,d}^{a}-\delta{}_{\,\,d}^{a}\mathring{L}{}_{\,\,b}^{c}\right),\nonumber \\
\left[\mathring{L}_{\,\,b}^{a},P_{c}\right] & = & -i\left(\delta_{\,\,c}^{a}P_{b}-\frac{1}{4}\delta_{\,\,b}^{a}P_{c}\right),\nonumber \\
\left[P_{a},P_{b}\right] & = & 0.
\end{eqnarray}

The group elements $g$ of the $SA\left(4,R\right)$ can be given
by exponential representation,
\begin{equation}
g\left(x,\mathring{\omega}\right)=e^{ix^{a}\left(x\right)P_{a}}e^{i\mathring{\omega}_{\,a}^{b}\left(x\right)\mathring{L}_{\,b}^{a}},
\end{equation}
where $x^{a}\left(x\right),\,\mathring{\omega}{}_{\,\,a}^{b}\left(x\right)$
are the real parameters. The MC structure equations of $SA\left(4,R\right)$
are found by using Eq. (\ref{eq: MC eq}) as follows,
\begin{eqnarray}
0 & = & d\Omega_{\,\,\,P}^{a}+\Omega_{\,\,\mathring{L}b}^{a}\wedge\Omega_{\,\,\,P}^{b}-\frac{1}{4}\Omega_{\,\,\mathring{L}}\wedge\Omega_{\,\,\,P}^{a},\nonumber \\
0 & = & d\Omega_{\,\,\mathring{L}b}^{a}+\Omega_{\,\,\mathring{L}c}^{a}\wedge\Omega_{\,\,\mathring{L}b}^{c},\label{eq: MC SA}
\end{eqnarray}
where the MC 1-forms $\Omega_{\,\,\,P}^{a}$ and $\Omega_{\,\,\mathring{L}b}^{a}$
correspond to translation and special-linear transformations. Considering
the methods given in \cite{bonanos2009,bonanos2010C,azcarraga2013A,azcarraga1995},
we can obtain the Maxwell extension of the special affine group as
$\mathcal{M}\mathcal{SA}\left(4,R\right)$ with the following non-zero
commutation rules by using the MC structure equations (\ref{eq: MC SA}),
\begin{eqnarray}
\left[\mathring{L}_{\,\,b}^{a},\mathring{L}_{\,\,d}^{c}\right] & = & i\left(\delta_{\,\,b}^{c}\mathring{L}_{\,\,d}^{a}-\delta{}_{\,\,d}^{a}\mathring{L}{}_{\,\,b}^{c}\right),\nonumber \\
\left[\mathring{L}_{\,\,b}^{a},P_{c}\right] & = & -i\left(\delta_{\,\,c}^{a}P_{b}-\frac{1}{4}\delta_{\,\,b}^{a}P_{c}\right),\nonumber \\
\left[P_{a},P_{b}\right] & = & iZ_{ab},\nonumber \\
\left[\mathring{L}_{\,\,b}^{a},Z_{cd}\right] & = & i\left(\delta_{\,\,d}^{a}Z_{bc}-\delta_{\,\,c}^{a}Z_{bd}+\frac{1}{2}\delta_{\,\,b}^{a}Z_{cd}\right).\label{eq: msa alg}
\end{eqnarray}

The non-linear realisation of $SA\left(4,R\right)$ can be found in
\cite{borisov1974,hamamoto1978}. To obtain a non-linear realisation
of the group $\mathcal{M}\mathcal{SA}\left(4,R\right)$ we will use
the techniques of coset realisation \cite{coleman1969,callan1969,salam1969A,salam1969B}.
We choose our coset as,
\begin{equation}
K(x,\theta)=\frac{\mathcal{M}\mathcal{SA}}{SL}=e^{ix\cdot P}e^{i\theta\cdot Z},
\end{equation}
where $x^{a},\,\theta^{ab}$ are the coset parameters and to find the
coset transformation we will use following formula,
\begin{equation}
g(a,\varepsilon,u)K(x,\theta)=K(x^{\prime},\theta^{\prime})h(\mathring{\omega}),
\end{equation}
where $\alpha,\,\epsilon,\,u$ are the real parameters for space-time
translations, tensorial translations and special linear group transformation
respectively. Thus the element of the stability subgroup takes the form
as $h(\mathring{\omega})=e^{i\mathring{\omega}{}_{\,\,a}^{b}\mathring{L}{}_{\,\,b}^{a}}$.
The infinitesimal transformation rules of the coset space parameters
under the action of the $\mathcal{MSA}(4,R)$ can be found as follows
by using the well-known Baker-Hausdorff-Campbell formula $e^{A}e^{B}=e^{A+B+\frac{1}{2}\left[A,B\right]}$,
\begin{align}
\delta x^{a} & =a^{a}+u_{\,\,c}^{a}x^{c}-\frac{1}{4}ux^{a},\label{eq:var_x}\\
\delta\theta^{ab} & =\varepsilon^{ab}+u_{\,\,\,c}^{[a|}\theta^{c|b]}-\frac{1}{2}u\theta^{ab}-\frac{1}{4}a^{[a}x^{b]},\label{eq:var_teta}\\
\mathring{\omega}_{\,\,b}^{a} & =u_{\,\,b}^{a}.
\end{align}

Now we can find the differential realisation of the corresponding
generators of $\mathcal{M}\mathcal{SA}\left(4,R\right)$ by comparing
the transformation rules of a scalar field $\Phi'\left(x^{a},\theta^{ab}\right)=\Phi\left(x^{a}-\delta x^{a},\theta^{ab}-\delta\theta^{ab}\right)$
and transformation of a scalar field under $\mathcal{M}\mathcal{SA}\left(4,R\right)$, defined as
\begin{equation}
\delta\Phi=i\left(a^{a}P_{a}+\varepsilon^{ab}Z_{ab}+u_{\,\,a}^{b}\mathring{L}_{\,\,b}^{a}\right).
\end{equation}
So we find the explicit form of the generators as follows,
\begin{align}
P_{a} & =i\left(\partial_{a}-\frac{1}{2}x^{b}\partial_{ab}\right),\nonumber \\
Z_{ab} & =i\partial_{ab},\nonumber \\
\mathring{L}_{\,\,b}^{a} & =i\left(x^{a}\partial_{b}-\frac{1}{4}\delta_{\,\,b}^{a}x^{c}\partial_{c}+2\theta^{ac}\partial_{bc}-\frac{1}{2}\delta_{\,\,b}^{a}\theta^{cd}\partial_{cd}\right),
\end{align}
where the derivatives are $\partial_{a}=\frac{\partial}{\partial x^{a}}$,
$\partial_{ab}=\frac{\partial}{\partial\theta^{ab}}$ and the self-consistency
of the algebra (\ref{eq: msa alg}) can be checked by Jacobi identities. 

\section{Gauging the Maxwell-special-affine algebra}
Let us now consider the gauge theory of $\mathcal{M}\mathcal{SA}\left(4,R\right)$.
We shall follow the same methods given in \cite{azcaraga2011,azcarraga2014,cebecioglu2014,cebecioglu2015}
in order to construct the gauge theory. For this purpose, we write
down a gauge field valued in the algebra as, 
\begin{equation}
\mathcal{A}=\mathcal{A}^{A}X_{A}=e^{a}P_{a}+B^{ab}Z_{ab}+\mathring{\omega}_{\,\,a}^{b}\mathring{L}_{\,\,b}^{a},
\end{equation}
where $X_{A}$ correspond to the generators of the algebra and the associated gauge fields $\mathcal{A}^{A}_{\mu}=\left(e^{a}_{\mu},B^{ab}_{\mu},\mathring{\omega}^{ab}_{\mu}\right)$ can be described by one-form fields $e^{a}$ $=e_{\mu}^{a}dx^{\mu}$, $B^{ab}=B_{\mu}^{ab}dx^{\mu}$
and $\mathring{\omega}{}_{\,\,b}^{a}=\mathring{\omega}{}_{\mu b}^{a}dx^{\mu}$
respectively. Using the Lie algebra valued parameter
$\zeta\left(x\right)$,
\begin{equation}
\zeta\left(x\right)=y^{a}\left(x\right)P_{a}+\varphi^{ab}(x)Z_{ab}+\lambda{}_{\,\,a}^{b}(x)\mathring{L}{}_{\,\,b}^{a},
\end{equation}
with the following gauge transformation,
\begin{equation}
\delta\mathcal{A}=-d\zeta-i\left[\mathcal{A},\zeta\right],\label{delta_amu}
\end{equation}
we get variations of the gauge fields, 
\begin{align}
\delta e^{a} & =\lambda_{\,\,b}^{a}e^{b}-\frac{1}{4}\lambda e^{a},\label{eq:vierbein_geuge_trans}\\
\delta B^{ab} & =\lambda_{\,\,\,c}^{[a|}B^{cb]}-\frac{1}{2}\lambda B^{ab}+\frac{1}{2}e^{[a}y^{b]},\\
\delta\mathring{\omega}_{\,\,b}^{a} & =0,
\end{align}
where $y^{a}\left(x\right)$, $\varphi^{ab}(x)$, and $\lambda{}_{\,\,a}^{b}(x)$
correspond to space-time translations, tensorial space, and the special
linear transformation parameters respectively. The curvature two-forms
of the associated gauge fields are given by,
\begin{equation}
\digamma=d\mathcal{A}+\frac{i}{2}\left[\mathcal{A},\mathcal{A}\right],
\end{equation}
where, 
\begin{equation}
\mathcal{\digamma}=F^{a}P_{a}+F^{ab}Z_{ab}+\mathcal{R}_{\,\,b}^{a}\mathring{L}{}_{\,\,a}^{b}.\label{f1}
\end{equation}

Thus we find the curvatures as follows,
\begin{eqnarray}
F^{a} & = & de^{a}+\mathring{\omega}_{\,\,b}^{a}\wedge e^{b}-\frac{1}{4}\mathring{\omega}\wedge e^{a}=\mathcal{\mathfrak{D}}e^{a},\\
F^{ab} & = & dB^{ab}+\mathring{\omega}_{\,\,\,c}^{[a|}\wedge B^{c|b]}-\frac{1}{2}\mathring{\omega}\wedge B^{ab}-\frac{1}{2}e^{a}\wedge e^{b},\label{eq: curv_Fab}\\
 & = & \mathcal{\mathfrak{D}}B^{ab}-\frac{1}{2}e^{a}\wedge e^{b},\\
\mathcal{R}_{\,\,b}^{a} & = & d\mathring{\omega}_{\,\,b}^{a}+\mathring{\omega}_{\,\,c}^{a}\wedge\mathring{\omega}_{\,\,b}^{c}=\mathcal{\mathfrak{D}}\mathring{\omega}_{\,\,b}^{a},
\end{eqnarray}
where the exterior covariant derivative with respect to $SL\left(4,R\right)$
defined as,
\begin{equation}
\label{cov_der_s}
\mathcal{\mathfrak{D}}\Phi:=\left[d+\mathring{\omega}+w\left(\Phi\right)\mathrm{Tr}\left(\mathring{\omega}\right)\right]\Phi;
\end{equation}
here $w$ behaves like the Weyl weight of corresponding field which occurs in the Weyl gauge theory as a result of scale transformations \cite{charap1974,cebecioglu2014}. The
infinitesimal gauge transformations of the curvatures under the related
symmetry can be found by,
\begin{equation}
\delta\mathcal{\digamma}=i\left[\zeta,\mathcal{\digamma}\right],
\end{equation}
and so we get,
\begin{eqnarray}
\delta F^{a} & = & -\mathcal{R}_{\,\,b}^{a}y^{b}+\frac{1}{4}\mathcal{R}y^{a}+\lambda_{\,\,b}^{a}F^{b}-\frac{1}{4}\lambda F^{a},\\
\delta F^{ab} & = & -\mathcal{R}_{\,\,\,c}^{[a|}\varphi^{c|b]}+\frac{1}{2}\mathcal{R}\varphi^{ab}+\lambda_{\,\,\,c}^{[a|}F^{c|b]}-\frac{1}{2}\lambda F^{ab}+\frac{1}{2}F^{[a}y^{b]},\\
\delta\mathcal{R}_{\,\,b}^{a} & = & \lambda_{\,\,c}^{a}\mathcal{R}_{\,\,b}^{c}-\lambda_{\,\,b}^{c}\mathcal{R}_{\,\,c}^{a}.
\end{eqnarray}

Taking covariant derivative of the curvatures, the generalized Bianchi
identities can be written as follows,
\begin{eqnarray}
\mathcal{\mathfrak{D}}F^{ab} & = & \mathcal{R}_{\,\,\,c}^{[a|}\wedge B^{c|b]}-\frac{1}{2}\mathcal{R}\wedge B^{ab}-\frac{1}{2}F^{[a}\wedge e^{b]},\nonumber \\
\mathcal{\mathfrak{D}}F^{a} & = & \mathcal{R}_{\,\,b}^{a}\wedge e^{b}-\frac{1}{4}\mathcal{R}\wedge e^{a},\nonumber \\
\mathcal{\mathfrak{D}}\mathcal{R}_{\,\,b}^{a} & = & 0.\label{eq:Bianchi}
\end{eqnarray}

\section{Topological gravity in four dimensions}
Now we can construct the gauge invariant Lagrangian 4-form under
local $SL(4,R)$ transformations with topological terms according
to the papers \cite{cebecioglu2015,mardones1991,mielke2012,mielke2008,sobreiro2011}.
We consider two possible topological 4-form. First, the gravitational
Pontryagin class topological term,
\begin{equation}
L_{Pontr}=\mathcal{R}_{\,\,b}^{a}\wedge\mathcal{R}_{\,\,a}^{b},
\end{equation}
where $L_{Pontr}$ is metric-free Lagrangian. The second is the
Nieh-Yan topological term \cite{nieh2007},
\begin{equation}
L_{NY}=\mathcal{R}_{ab}\wedge e^{a}\wedge e^{b}+F^{a}\wedge F_{a},
\end{equation}
where $F^{a}$ is the torsion tensor. Here the NY term is needed to
introduce a metric tensor $g_{ab}$ to raise and lower the indices
for example $F_{a}=g_{ab}F^{b}$ \cite{mielke2012}. For our Lagrangian
we define a shifted curvature as,
\begin{equation}
\mathcal{Y}{}_{\,\,b}^{a}:=\mathcal{R}_{\,\,b}^{a}-\mu F{}_{\,\,b}^{a},\label{eq: shifted curvature}
\end{equation}
 and one can find its infinitesimal gauge transformation under $SL\left(4,R\right)$
symmetry as,
\begin{equation}
\delta\mathcal{Y}_{\,\,b}^{a}=\lambda_{\,\,c}^{a}\mathcal{Y}_{\,\,b}^{c}-\lambda_{\,\,b}^{c}\mathcal{Y}_{\,\,c}^{a}.
\end{equation}

After these preliminaries we can write the gravity action that is invariant
under local $SL\left(4,R\right)$ symmetry as follows,
\begin{equation}
S=\frac{1}{2\varkappa}\int\mathcal{Y}{}_{\,\,b}^{a}\wedge\mathcal{Y}{}_{\,\,a}^{b}+\frac{1}{\rho}\int F^{a}\wedge F_{a},\label{action-1}
\end{equation}
where $\varkappa=8\pi G$ is the Einstein gravitational constant and $\rho$
is an arbitrary constant. One can also show the action Eq. (\ref{action-1})
is diffeomorphism invariant. If we decompose the action (\ref{action-1}),
we get,
\begin{eqnarray}
S & = & \frac{1}{2\varkappa}\int\left(\mathcal{R}_{\,\,b}^{a}-\mu F{}_{\,\,b}^{a}\right)\wedge\left(\mathcal{R}_{\,\,a}^{b}-\mu F{}_{\,\,a}^{b}\right)+\frac{2\chi}{\rho}F^{a}\wedge F_{a}\nonumber \\
& = & \frac{1}{2\varkappa}\int
\mathcal{R}_{\,\,b}^{a}\wedge\mathcal{R}_{\,\,a}^{b}+\left(\mu\mathcal{R}_{\,\,b}^{a}\wedge e^{b}\wedge e_{a}+\frac{2\chi}{\rho}F^{a}\wedge F_{a}\right)\nonumber\\
& &
+\mu^{2}\mathcal{\mathfrak{D}}B_{\,\,b}^{a}\wedge\mathcal{\mathfrak{D}}B_{\,\,a}^{b}-2\mu\mathcal{R}_{\,\,b}^{a}\wedge\mathcal{\mathfrak{D}}B_{\,\,a}^{b}-\frac{\mu^{2}}{2}\mathcal{\mathfrak{D}}B_{\,\,b}^{a}\wedge e^{b}\wedge e_{a}\nonumber\\
& &
+\frac{\mu^{2}}{4}e^{a}\wedge e_{b}\wedge e^{b}\wedge e_{a},
\end{eqnarray}
where, in the last result, the first term is the Pontrjagin 4-form,
the second one can be seen as a NY term, and the others are the contributions
come from Maxwell symmetry. Let us derive the field equations. The
variation of Eq.(\ref{action-1}) with respect to gauge field $\mathring{\omega}_{\,\,b}^{a}$
we get, 
\begin{equation}
\mathcal{\mathfrak{D}}\mathcal{Y}_{\,\,b}^{a}-\mu\left[B,\mathcal{Y}\right]_{\,\,b}^{a}+\frac{\chi}{\rho}\left[e,F\right]_{\,\,b}^{a}=0,\label{B1}
\end{equation}
where the expression $\left[B,\mathcal{Y}\right]{}_{\,\,b}^{a}$ $=\left(B_{\,\,c}^{a}\wedge\mathcal{Y}{}_{\,\,b}^{c}-\mathcal{Y}{}_{\,\,c}^{a}\wedge B{}_{\,\,b}^{c}\right)$,
and similarly we have $\left[e,F\right]{}_{\,\,b}^{a}=(e^{a}\wedge F_{b}-F^{a}\wedge e_{b}).$
Taking the $e^{a}$ and $e_{a}$ variation of the action respectively,
\begin{eqnarray}
e_{b}\wedge\mathcal{Y}{}_{\,\,a}^{b}+\left(\frac{2\varkappa}{\mu\rho}\right)\mathcal{\mathfrak{D}}F_{a} & = & 0,\label{B2}\\
\mathcal{Y}{}_{\,\,b}^{a}\wedge e^{b}-\left(\frac{2\varkappa}{\mu\rho}\right)\mathcal{\mathfrak{D}}F^{a} & = & 0,
\end{eqnarray}
and finally variation with respect to $B_{\,\,b}^{a}$ field,
\begin{equation}
\mathcal{\mathfrak{D}Y}{}_{\,\,b}^{a}=0.\label{B3}
\end{equation}
These are the equations of motions of the action given in Eq. (\ref{action-1})
under local $SL\left(4,R\right)$. One can also obtain following constraint
equation by substituting Eq.(\ref{B3}) into Eq.(\ref{B1}) and taking
exterior covariant derivative and using the Eq.(\ref{B2}), 
\begin{equation}
\mathcal{Y}_{\,\,c}^{a}\wedge F_{\,\,b}^{c}-F_{\,\,c}^{a}\wedge\mathcal{Y}_{\,\,b}^{c}=0.\label{constraint}
\end{equation}

This constraint equation can be solved by the selection as $\mathcal{Y}_{\,\,b}^{a}=0$
or $\mathcal{Y}_{\,\,b}^{a}=\alpha F_{\,\,b}^{a}$, so we get the
relation between two curvatures as $\mathcal{R}_{\,\,b}^{a}\thicksim F_{\,\,b}^{a}$.
On the other hand, if we select a special solution for the constraint
Eq.(\ref{constraint}) as,
\begin{equation}
\mathcal{Y}_{\,\,b}^{a}=-2\mu F_{\,\,b}^{a}+\frac{1}{2}\delta_{\,\,b}^{a}\mathcal{R},\label{eq: special sol}
\end{equation}
and by the help of comparison between the definition of $\mathcal{Y}_{\,\,b}^{a}$
given in Eq. (\ref{eq: shifted curvature}) and the special solution
Eq. (\ref{eq: special sol}), we obtain following equation, 
\begin{eqnarray}
\mathcal{R}_{\,\,b}^{a}-\frac{1}{2}\delta_{\,\,b}^{a}\mathcal{R} & = & -\mu F_{\,\,b}^{a}.\label{eq:solution_1}
\end{eqnarray}

Thus the Eq. (\ref{eq: special sol}) takes following form,
\begin{equation}
\mathcal{Y}_{\,\,b}^{a}=2\mathcal{R}_{\,\,b}^{a}-\frac{1}{2}\delta_{\,\,b}^{a}\mathcal{R}.\label{eq: solution_2}
\end{equation}

From these results, we get the constraint as $\mathfrak{D}F_{\,\,\,b}^{a}=0$
and taking account of Eqs. (\ref{eq: special sol}, \ref{eq: solution_2})
and equations of motions we get the generalized Bianchi identities
of $\mathcal{M}\mathcal{SA}\left(4,R\right)$ symmetry given in Eq.
(\ref{eq:Bianchi}) under $\left(\frac{\varkappa}{\mu\rho}\right)\rightarrow1$
limit. Furthermore, the left hand side of the Eq. (\ref{eq:solution_1})
is similar to the form of Einstein tensor. If we take the Eq. (\ref{eq:solution_1})
as the Einstein field equation then the tensor $F_{\,\,b}^{a}$ corresponds
to energy momentum tensor.

\section{Conclusion}

In this letter, we have presented a tensor extension of the special-affine $SA\left(4,R\right)$ group by adding six abelian tensorial generators. Then the nonlinear realization of the Maxwell-special-affine group $\mathcal{MSA}\left(4,R\right)$ was constructed and the explicit form of the corresponding generators was found. After that, we constructed the gauge theory of $\mathcal{MSA}\left(4,R\right)$ and wrote down a gravity action by using the Pontryagin and the NY invariants. We also observed that the equations of motion go to the generalized Bianchi identities in a certain condition. Moreover, we obtained the form of Einstein field equation as a special solution of the equations of motion
and from this result, we can say that background energy leads to space-time curvature. Thus we obtained a restricted version of the results given in the paper \cite{cebecioglu2015} because of using $\mathcal{MSA}\left(4,R\right)$ instead of $\mathcal{MGA}\left(4,R\right)$.

The affine symmetry affords us the most general possible transformations within a
spacetime framework by using the general linear connection $\Gamma^{\alpha}_{\mu\nu}$ which has 64 independent fields\cite{hehl1977}. So there are widespread usage areas for affine symmetry from particle physics to gravitation  \cite{neeman1979B,neemann1988B,hehl1995,tresguerres2000,hehl1978,hehl1991,leclerc2006,julia1998}.
For instance, according to the papers\cite{neeman1979A,neeman1979B,neemann1988A,neemann1988B,hehl1995,lopez-pinto1995}, the gauge theory of affine group could play an important role for solving the renormalizability or unitarity problems on quantum gravity by means of the general linear connection which contains additional degrees of freedom. Also, the paper \cite{borzeszkowski1996} demonstrated that the pure affine field theory developed by Einstein \cite{einstein1955} and Schr\"{o}dinger \cite{schr=0000F6dinger1950}, known as the Einstein-Schr\"{o}dinger theory, provides the unification
of quantum theory and Einstein's general theory of relativity and
this theory can be quantized by the rules of canonical quantization, but the author says that this quantization is physically meaningless. These studies demonstrate us the importance of the affine symmetry. So the Maxwell extension of the affine symmetry proposes an extended framework for the mentioned problems. Also, metric-affine gravity theories based on $\mathcal{MA}\left(4,R\right)$ and $\mathcal{MSA}\left(4,R\right)$
are in progress.

\end{document}